# Complexity in the Wake of Artificial Intelligence

Theodore Modis

Growth Dynamics, Via Selva 8, 6900 Massagno, Lugano, Switzerland.

**Abstract:** This study aims to evaluate quantitatively (albeit in arbitrary units) the evolution of complexity of the human system since the domestication of fire. This is made possible by studying the timing of the 14 most important milestones – breaks in historical perspective – in the evolution of humans. AI is considered here as the latest such milestone with importance comparable to that of the Internet. The complexity is modeled to have evolved along a bell-shaped curve, reaching a maximum around our times, and soon entering a declining trajectory. According to this curve, the next evolutionary milestone of comparable importance is expected around 2050-2052 and should add less complexity than AI but more than the milestone grouping together nuclear energy, DNA, and the transistor. The peak of the complexity curve coincides squarely with the life span of the baby boomers. The peak in the rate of growth of the world population precedes the complexity peak by 25 years, which is about the time it takes a young man or woman before they are able to add complexity to the human system in a significant way. It is in society's interest to flatten the complexity bell-shaped curve to whatever extent this is possible in order enjoy complexity longer.

**Keywords:** complexity, entropy, AI, logistic growth, evolutionary milestones, baby boom, world population

## 1. Introduction

Entropy and complexity are subjects that have enjoyed enormous attention in the scientific literature. There have been many definitions for entropy and even more for complexity. John Horgan in his June 1995 *Scientific American* editorial entitled "From complexity to perplexity" mentioned a list of 31 definitions of complexity.[Horgan, 1995] But for the purposes of this work we will define entropy and complexity as follows. Intuitively, as a measure of disorder, and how difficult it is to describe, respectively. More rigorously, with information-related definitions, entropy as the information content,[Shannon, 1948] and complexity as the capacity to incorporate information, in line with the thinking of Gell-Mann [Gell-Mann, 1994] and Simon [Simon, 1996].

It is worth mentioning that our definition of complexity corresponds to what Pier Luigi Gentili calls Descriptive Complexity, which is a combination of Effective Complexity and Shannon Entropy.[Gentili, 2021] Effective Complexity is related to Algorithmic Information Content and Kolmogorov Complexity.[Gell-Mann and Lloyd, 1996]

Life and particularly humans have a profound impact on entropy. They decrease it locally by creating and maintaining highly ordered and complex structures. But the overall entropy $S$ always increases in an isolated system according to the 2$^{nd}$ law of thermodynamics, namely $\Delta S \geq 0$. In fact, the more humans decrease entropy locally, the more entropy will increase elsewhere. Humans and life in general evolve and thrive by increasing the overall entropy, which grows monotonically following some kind of S-shaped trajectory.

Complexity also grows in the beginning but eventually, it declines because it follows overall some kind of bell-shaped trajectory. The concept of complexity eventually decreasing has been popularized by world-renowned scientists. In his bestselling book *The Big Picture: On the Origins of Life, Meaning, and the Universe Itself* theoretical physicist Sean Carroll argues that complexity is related to entropy and that "complexity is about to begin declining."[Carroll, 2016] In his book *The Quark and the Jaguar: Adventures in the Simple and the Complex*, Nobel laureate Murray Gell-Mann argues that there is a trade-off between entropy and complexity, and that as entropy increases, complexity may increase only up to a certain point beyond which the system becomes too disordered to sustain its complexity.[Gell-Mann,1994]

In a way that reminds us of cliometrics (econometric history) this work updates, expands, and explores further previously published works. It studies evolutionary milestones – breaks in historical perspective – in order to quantify the evolution of complexity. In 2002 a similar study analyzed 28 such milestones beginning with the Big Bang and ending with Internet/sequencing of the human genome in 1995.[Modis, 2002] That study considered complexity to be intricately linked to change. In the interest of the reader, some of the discussion from that study is reproduced below, beginning with a quotation: "Complexity increases both when the rate of change increases and when the amount of things that are changing around us increase. Our task then becomes to quantify complexity, as it evolved over time, in an objective, scientific way and therefore defensible way. Also to determine the law that best describes complexity's evolution over time, and then to forecast its future trajectory. This will throw light onto what one may reasonably expect as the future rate at which change will appear in society."

We have seen much literature and extensive preoccupation of "hard" and "less hard" scientists with the subject of complexity. Yet we have neither a satisfactory definition for it, nor a practical way to measure it. The term complexity remains today vague and unscientific. In his best-selling book *Out of Control,* Kevin Kelly concludes:

“How do we know one thing or process is more complex than another? Is a cucumber more complex than a Cadillac? Is a meadow more complex than a mammal brain? Is a zebra more complex than a national economy? I am aware of three or four mathematical definitions for complexity, none of them broadly useful in answering the type of questions I just asked. We are so ignorant of complexity that we haven’t yet asked the right question about what it is.” [Kelly, 1994]

But let us look more closely at some of the things that we do know about complexity:

- It is generally accepted that complexity increases with evolution. This becomes obvious when we compare the structure of advanced creatures (animals, humans) to primitive life forms (worms, bacteria).
- It is also known that evolutionary change is not gradual but proceeds by jerks. In 1972 Niles Eldredge and Stephen Jay Gould introduced the term “Punctuated Equilibria”: long periods of changelessness or stasis – equilibrium – interrupted by sudden and dramatic brief periods of rapid change – punctuations.[Eldredge and Gould, 1972]

These two facts taken together imply that complexity itself must grow in a stepladder fashion, at least on a macroscopic scale.

We also know that:

- Complexity begets complexity. A complex organism creates a niche for more complexity around it; thus complexity is a positive feedback loop amplifying itself. In other words, complexity has the ability to “multiply” like a pair of rabbits in a meadow.
- Complexity links to connectivity. A network's complexity increases as the number of connections between its nodes increases, and this enables the network to evolve. But you can have too much of a good thing. Beyond a certain level of linking density, continued connectivity decreases the adaptability of the system as a whole. Kauffman calls it “complexity catastrophe”: an overly linked system is as debilitating as a mob of uncoordinated loners.[Kauffman, 1995]

These two facts argue for a process similar to growth in competition. Complexity is endowed with a multiplication capability but its growth is capped and that necessitates some kind of a selection mechanism. Alternatively, the competitive nature of complexity’s growth can be sought in its intimate relationship with evolution, namely that entropy reflects the accumulation of complexity.[Modis, 2022] One way or another, it is reasonable to expect that complexity follows a bell-shaped pattern as it grows.

Most teachers of biology, biochemistry, and geology at some time or another present to their students a list of major events in the history of life. The dates they mention invariably reflect milestones of punctuated equilibrium. Physicists tend to produce a different list of dates stretching over another time period with emphasis mostly on the early Universe. All milestones constitute turning points, beyond which the world is no longer the same as before.

Such lists constitute data sets that may be plagued by numerical uncertainties and personal biases depending on the investigator's knowledge and specialty. Nevertheless, the events listed in them are “significant” because an investigator has singled them out as such among many other events. Consequently, they constitute milestones that can in principle be used for the study of complexity’s evolution over time. However, in practice, there are some formidable difficulties in producing a data set of turning points that cover the entire period of time (13.8 billion years).

The study of 2002 mentioned earlier [Modis, 2002] made the bold hypothesis that a law has been in effect from the very beginning. This was not an arbitrary decision. It followed a first look at an early compilation of milestones. In any case, confrontation with the final data set is the ultimate judge. The scientific method – as defined by experimental physicists – says: Following an observation (or hunch), make a hypothesis, and see if it can be verified by real data.

The conclusion of the 2002 study was that the evolution of complexity in our world was approaching its maximum and should begin declining in the not-so-distant future tracing out a bell-shaped pattern. A decline in complexity had been forecasted for the 29$^{th}$ milestone at around 2033 and one possible candidate for the 29$^{th}$ milestone was Artificial Intelligence (AI).[Modis, 2022]

Today the author considers the emergence of AI in 2023 to be a milestone comparable in importance to the Internet, the transistor, nuclear energy, the printing press, etc. Using the same methodology the author updates and confronts the results reported in the two previous publications. While the approach is the same, there are two improvements. First, the time window studied is now restricted to only the human system – 14 evolutionary milestones – beginning with the domestication of fire 700,000 years ago and ending with AI in 2023. This restriction is done mainly because the complexity of the early-universe milestones is so small (in relative terms) that does not influence the determination of the final bell-shaped curve. In fact, all milestones before **Renaissance** (printing press, etc.) follow a purely exponential trend and impact minimally the determination of the bell-shaped curve. But also because the whole approach becomes more coherent and defensible if we focus only on the human system. Second, more new data were added because the majority of the data sets considered in the old studies did not extend into the 20$^{th}$ century.

To offset the inherent subjectivity in choosing milestones an effort has been made to combine inputs from world-renowned scientists (emailed more than 100 Nobel laureates in the sciences) and other reputable sources. The data are described in Section 2, the analysis in Section 3, and a discussion of the robustness of the results in Section 4. The results are discussed in Section 5, where we also see complexity linked to the population. There is a curious overlap of the ill-understood baby boom with the peak of the complexity curve. But also the peak in the rate of growth of the global population resembles and precedes the complexity peak by 25 years. Finally, there are some general conclusions in Section 6.

In the interest of the reader, a fair amount of text from previous publications [Modis, 2002, 2022, 2024] has been repeated/adapted here in the analysis and discussion sections. It contributes to a smoother reading and spares the reader the effort of searching the original publications for consultation.

## 2. The data

Three data sets with milestones from the original [Modis,2002] study have also been used here, namely:

- Carl Sagan's *Cosmic Calendar* [Sagan, 1986]
- a set of 25 milestones provided by Paul D. Boyer, biochemist, Nobel Prize 1997
- a set of 25 milestones provided together by the author and Eric L. Schwartz, professor of Cognitive and Neural Systems at Boston University.
- To the above have now been added 25 milestones furnished by ChatGPT, carefully verified and edited by the author.

Finally, there have been contributions for specific milestones by the following world-renowned scientists. They saw (and tacitly approved) the list in Appendix A, each one of them suggesting only one or two additional milestones that he deemed should also be included:

- Sir John Ernest Walker biochemist, Nobel Prize 1997
- Sir Peter John Ratcliffe, physician-scientist, Nobel Prize 2019
- Pierre Darriulat, Research Director at CERN, 1987 – 1994
- Athanasios G. Konstantopoulos, chemical engineer, professor at the Aristotle University of Thessaloniki, *Chevalier de l' Ordre national de la Légion d' Honneur*

The sum of all milestones thus compiled came to a total of 128.

Beginning with the domestication of fire, a list of the most important milestones – some of them obviously more important than others – is shown in Appendix A. In bold are shown the ***major* milestones** defined as events mentioned at least twice in the entire dataset.

It should be pointed out that the assigned dates reflect the date a milestone's impact began being felt significantly in society and not the date on which the phenomenon/discovery in question was first documented.

### 2.1 All milestones

In Figure 1, we see the 128 milestones plotted in a histogram with geometrically increasing time bins as we go back in time in order to accommodate the long time horizon. Besides the crowding of milestones in recent times, it is evident that there is clustering of these milestones. The peak of each cluster, determined by the weighted average of the dates of the milestones in the cluster, is used to define the date in a final set of 14 thus called "canonical" milestones. This is why some events may appear dated somewhat off, e.g., WWI, which belongs in the most dispersed cluster consisting of 19 milestones – canonical milestone No. 11 – appears to be positioned at 104 years before 2000.

The breadth of each cluster is used to calculate the error on the date of the peak (the mean) as the mean absolute error with respect to the mean. This error gets then propagated to an error on the value of the complexity calculated. As indicated in the graph the most dispersed cluster milestone No. 11 has the biggest error. Bins with only one entry have been assigned half the bin width as an approximation for full width half max (FWHM) in the calculation of the error. These are only statistical errors stemming from the methodology used. No attempt is made to estimate systematic errors.

The *major* milestones are highlighted in bold in Appendix A. They constitute the most important milestone(s) in a given cluster and their timing (in bold) is generally close to the peak of the cluster.

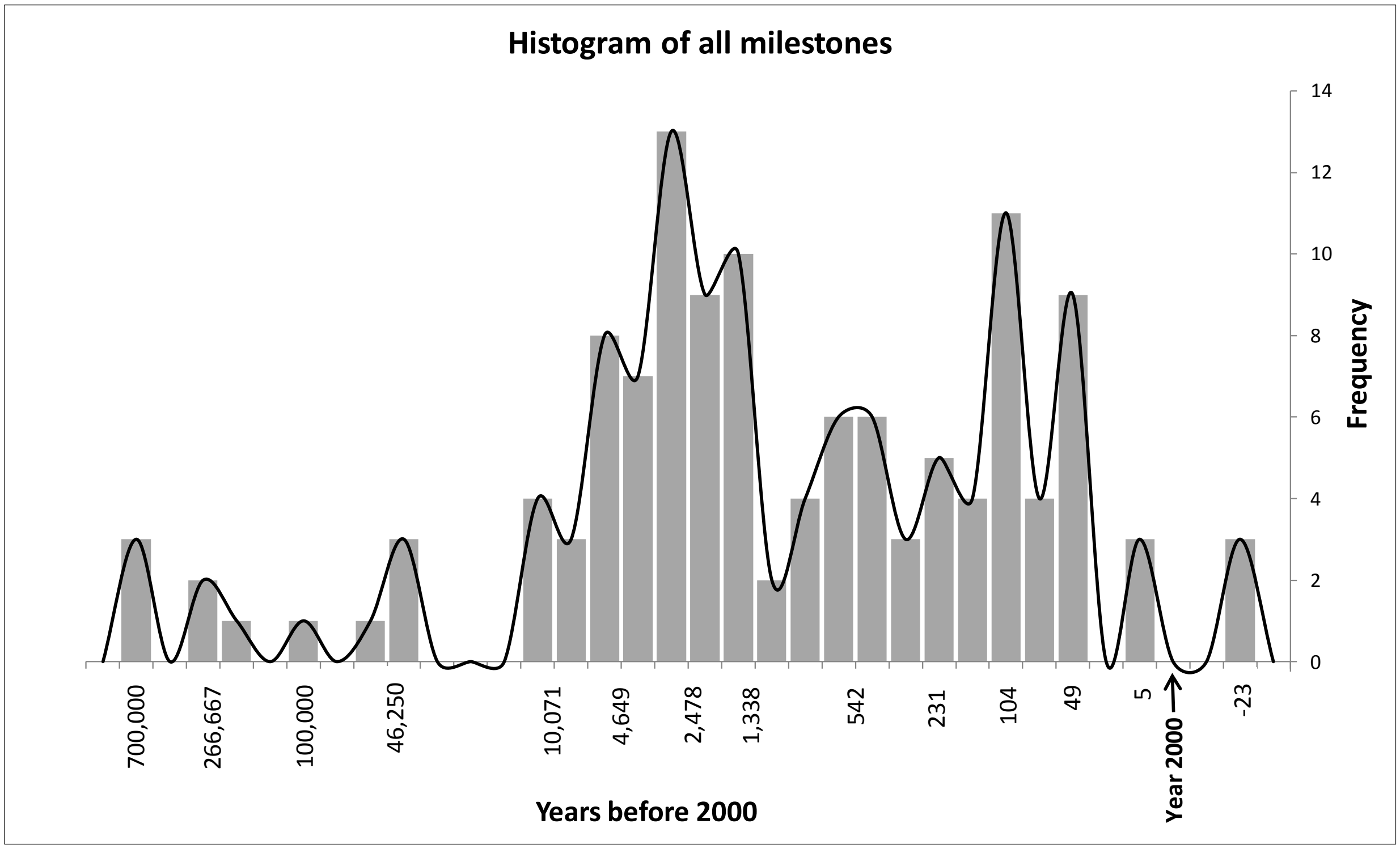

Figure 1. A histogram of the 128 milestones with geometrically increasing time bins as we go back in time. The thin black line is superimposed to outline the peaks that define the dates of the "canonical" milestone set. On the horizontal axis, we read the dates of these peaks determined as described in the text. The breadth of each cluster helps define the error on each date.

### 2.2 The major milestones

In Figure 2 we see the 56 *major* milestones, as defined earlier, plotted in a histogram with geometrically increasing time bins as we go back in time in order to accommodate the long time horizon. Clustering is now rudimentary, there are only a few entries per bin and there is no overlap between adjacent milestones. If there is only one milestone in the bin, the date assigned is that of the milestone. If there are more than one milestones in the bin, the date assigned is the average date. The error on each date is calculated as the mean absolute error with respect to the average. These errors then get propagated to errors in the values of the complexity calculated. Once again no attempt is made to take into account systematic errors.

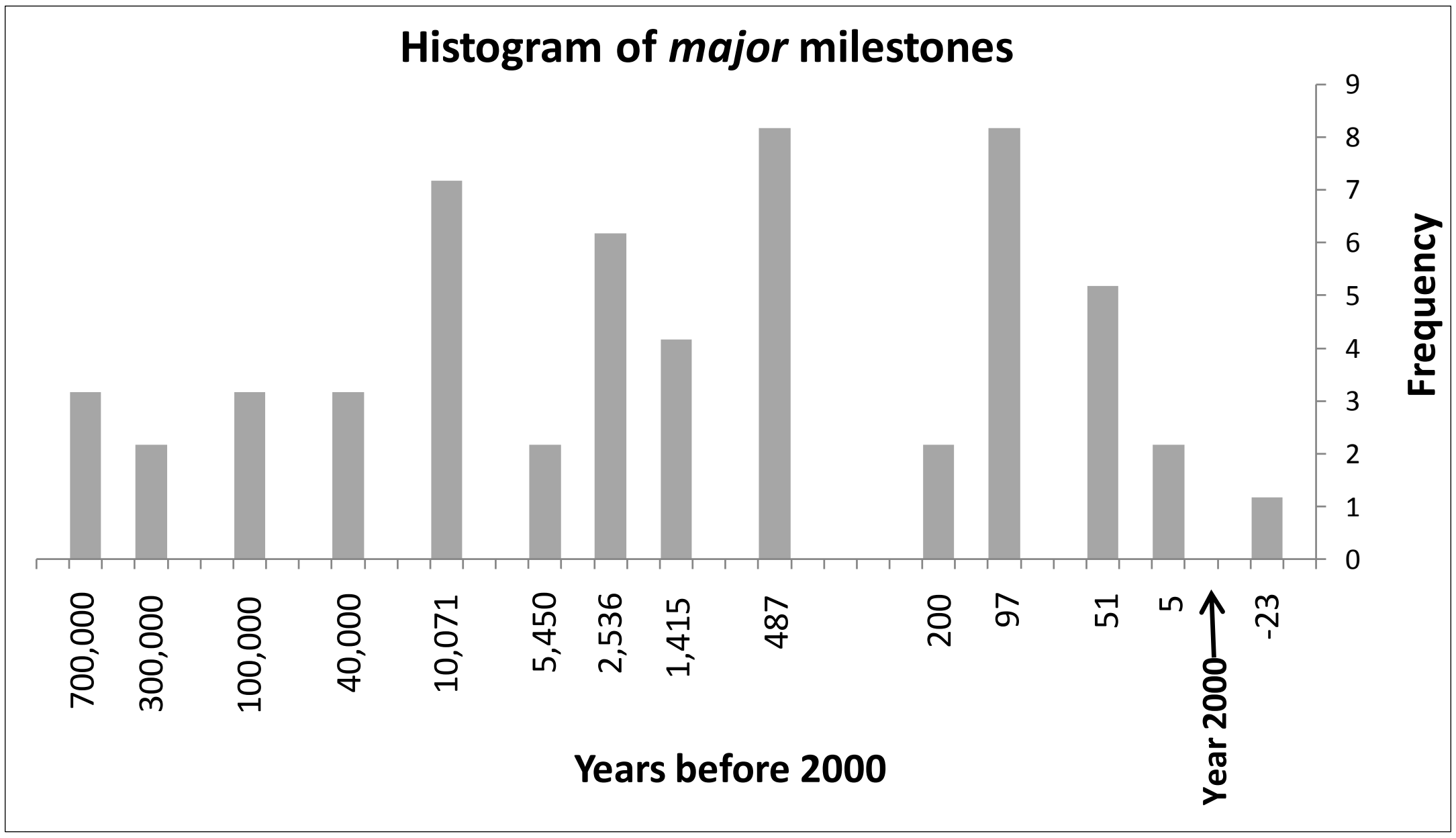

Figure 2. A histogram of 56 *major* milestones with geometrically increasing time bins as we go back in time. On the horizontal axis, we read the average date of the milestones in each bin. There is no overlap between adjacent milestones.

## 3. The Analysis

It is easy to quantify the complexity of a simple system like a fair dice.[Modis, 2024] But it seems hopelessly unrealistic to quantify complexity for humans and their environment in absolute terms. A more realistic endeavor is to quantify complexity in *relative* terms as was done in the original study.[Modis, 2002] To facilitate the reader we reproduce below the steps involved.

The complexity associated with an evolutionary canonical milestone is quantified according to the event's importance. *Importance* can be defined as *the change in complexity multiplied by the time duration to the next milestone*. This definition has been derived in the classical physics tradition: you start with a magnitude (in our case *Importance*), you put an equal sign next to it, and then you proceed to list in the numerator whatever the quantity in question is proportional to, and in the denominator whatever it is inversely proportional to, keeping track of possible exponents and multiplicative constants. It is intuitively obvious that for a milestone *Importance* is linearly proportional to the amount of complexity added by the milestone, and also linearly proportional to how long the system survives unchanged following the milestone. The greater the complexity jump at a given milestone, or the longer the ensuing stasis, the greater the milestone's importance will be:

$$Importance \propto Complexity \times Duration$$

Despite the fact that in the set of 128 milestones, there are milestones of lesser importance, the 14 *canonical* milestones are considered to be of *utmost* importance, and to that extent, we can approximate them as being of *equal* importance.

Following each milestone the complexity of the system increases by a certain amount. At the next milestone, there is another increase in complexity. Assuming that milestones are approximately of equal importance, and according to the above definition of importance, we can conclude that the increase in complexity $\Delta C_i$ associated with milestone *i* of importance *I* will be inversely proportional to the time period to the next milestone. We can thus quantify the complexity of milestone *i* as follows:

$$\Delta C_i = \frac{I}{\Delta T_i} \qquad (1)$$

where *I* the importance (in arbitrary units) is the same for all canonical milestones, and $\Delta T_i$ is the time period between milestone *i* and milestone *i*+1.

Equation (1) provides a quantitative *relative* determination of the complexity contributed by each canonical milestone to the system. If milestones become progressively crowded together with time, their complexity is expected to become progressively larger.

We saw earlier that complexity constitutes a positive feedback loop amplifying itself, and thus has the ability to "multiply." But only up to a point because too much complexity emulates simplicity. Entropy, which results from the accumulation of complexity, grows exponentially in the beginning and continues growing monotonically according to the 2$^{nd}$ law of thermodynamics. But it eventually approaches a maximum – a state of complete disorder – rather slowly, i.e. asymptotically. Consequently, entropy's trajectory follows some kind of an S-shaped curve, and goes through an inflection point around the middle, at a time when complexity goes over a maximum before beginning decreasing. A large-scale example is the entire Universe. Entropy began increasing at the beginning of the Universe with the Big Bang, when the Universe is thought to have been a smooth, hot, rapidly expanding plasma and rather orderly; a state with low entropy and low information content. Entropy will reach a maximum at the end of the Universe, which in a prevailing view will be a state of heat death, after black holes have evaporated and the acceleration of the Universe has dispersed all energy and particles uniformly everywhere [Carroll, 2010]. The information content of this final state of maximal disorder (everything being everywhere), namely the knowledge of the precise position and velocity of every particle in it will also reach a maximum. Entropy's trajectory grew rapidly during the early Universe. As the Universe's expansion accelerated, entropy's growth accelerated. Its trajectory followed a rapidly rising exponential-like growth pattern. At the other end, heat death, entropy will grow slowly to asymptotically reach the ceiling of its final maximum [Patel, 2019]. It will most likely happen along another exponential-like pattern. It follows that the overall trajectory of entropy will trace some kind of an S-shaped curve with an inflection point somewhere around the middle.

It is reasonable that a logistic function – a natural-growth curve – could be suitable to describe the data of the accumulated complexity. But the time frame considered by this analysis is vast and the milestones crowd together progressively more and more in recent times. The logistic pattern as a function of time cannot describe this growth trajectory adequately. A Euclidean (linear) conception of time is not appropriate for such an evolution. A more suitable time variable is the sequential milestone number, which represents some type of a "logistic" time scale, which is nonlinear with time stretching out both as $t \rightarrow \infty$ and as $t \rightarrow -\infty$.

Given that our data depict a *rate* of growth – i.e. complexity change per milestone – we expect their trend to follow the time derivative of the logistic function, i.e. the logistic life cycle. We therefore fit to the expression:

$$f'(x) = \frac{M\alpha}{(1+e^{-\alpha(x-x_0)})(1+e^{\alpha(x-x_0)})} \qquad (2)$$

where $M$, $\alpha$, and $x_o$ are constants, and $x$ is the sequential milestone number. The logistic life cycle is the first derivative of the familiar logistic function:

$$f(x) = \frac{M}{(1+e^{-\alpha(x-x_0)})} \qquad (3)$$

### 3.1 All milestones

Fitting Equation (2) to the data of all milestones as listed in the table of Appendix A yields the picture in Figure 3. We see the complexity of each canonical milestone with its error and the fitted logistic life cycle (thick gray line.) Table I shows the particular details of the fit. The goodness of the fit has been evaluated with a Graphical Analysis of Residuals, plotting the fit vs. the data with a trend line:

$Y = 0.9809 * X - 0.00019$ with $R^2 = 0.9853$

which indicates good accuracy ($R^2 \approx 1$), no systematic bias (intercept ≈ 0), and no data-dependent bias (slope ≈ 1).

**Table I - Fit Results - All milestones**

| **Function fitted** | | | | **Goodness of Fit** | | |
|---|---|---|---|---|---|---|
| $\frac{M\alpha}{(1+e^{-\alpha(x-x_0)})(1+e^{\alpha(x-x_0)})}$ | $\alpha$ | $M$ | $x_o$ | $R^2$ | **Slope** | **Intercept** |
| | 0.7907 | 0.1945 | 13.75 | 0.985 | 0.981 | -0.0002 |

The overall trajectory of the complexity indicates that it is presently at a maximum. It yields forecasts for the complexity of future milestones. Using the definition of *Importance* – in conjunction with the equi-importance assumption – we can then derive explicit dates for the future milestones, see Table III below.

In Figure 3 we see data only for 13 out of the 14 canonical milestones studied. The complexity for the 14$^{th}$ milestone (AI) cannot yet be calculated because we do not know how far in the future is the 15$^{th}$ milestone. The time difference between the 14$^{th}$ and the 15$^{th}$ milestones will define the complexity of the 14$^{th}$ milestone.

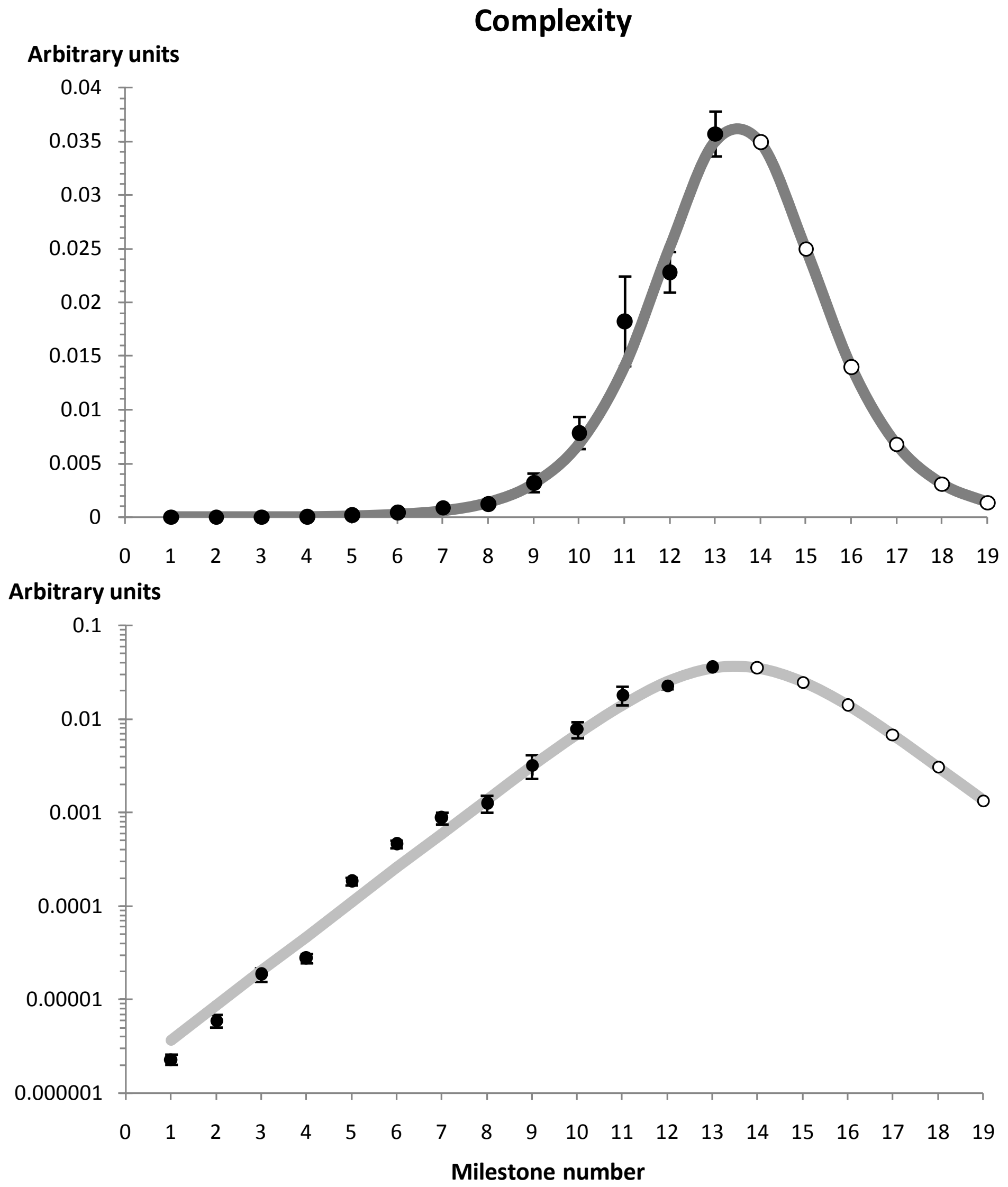

Figure 3. The thick gray line is a logistic life-cycle fit to the data of the first thirteen canonical milestones. The vertical axis depicts the change in complexity (with a logarithmic scale in the lower graph.) The little white circles on the extrapolated trend indicate the expected complexity of future milestones; the first one – Number 14 – refers to AI.

### 3.2 Major milestones

We repeat now the above exercise only for the *major* milestones, as they were defined earlier. We obtain again a good fit with results not significantly differentiated, see Table II. In Figure 4 we see again data only for 13 out of the 14 milestones studied. The complexity of the 14$^{th}$ milestone (AI) will be fixed when the timing of the 15$^{th}$ milestone becomes known.

The fitted trajectory of the complexity indicates again that we are presently at a maximum. The forecasts for the complexity of future milestones are very similar to those of Section 3.1, see Table III below.

**Table II - Fit Results – Major milestones**

| Function fitted | | | | Goodness of Fit | | |
|---|---|---|---|---|---|---|
| $\frac{M\alpha}{(1+e^{-\alpha(x-x_0)})(1+e^{\alpha(x-x_0)})}$ | $\alpha$ | $M$ | $x_o$ | $R^2$ | Slope | Intercept |
| | 0.8119 | 0.1849 | 13.71 | 0.980 | 1.0174 | 0.00049 |

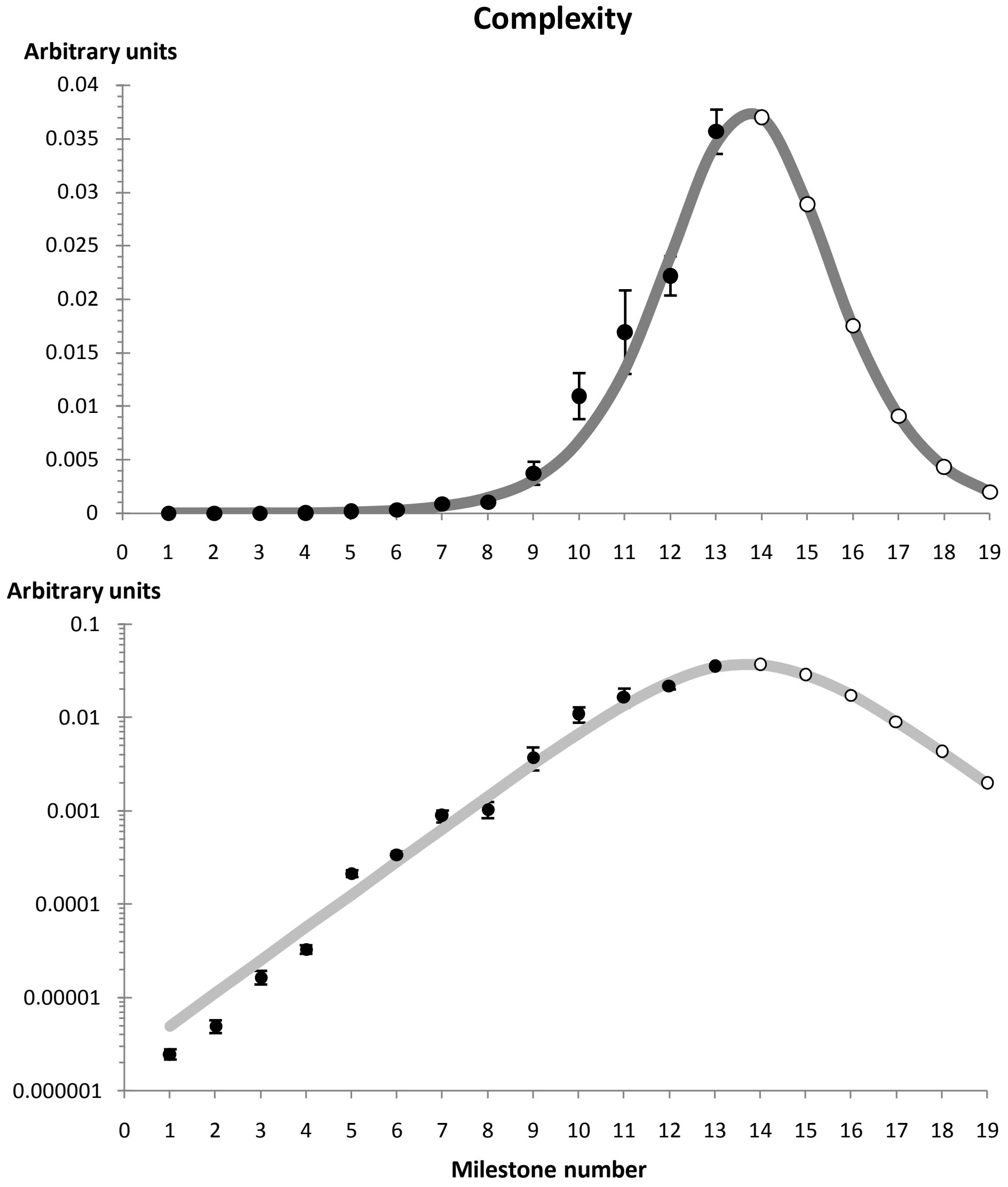

Figure 4. The thick gray line is a logistic life-cycle fit to the data of the first thirteen major milestones. The vertical axis depicts the change in complexity (with a logarithmic scale in the lower graph.) The little white circles on the extrapolated trend indicate the expected complexity of future milestones; the first one – Number 14 – refers to AI.

The little open circles in Figures 3 and 4 forecast the complexity values for AI and other future milestones. The forecasted complexity of future milestones can be translated to dates using Equation (1). Table III gives forecasts for the complexity contribution of AI and the future five milestones, as well as the dates on which they should expected.

**Table III – Complexity Forecasts***

| Milestone number | All milestones | Year | Major milestones | Year |
|---|---|---|---|---|
| 14 (AI) | 0.0350 | 2023 | 0.0363 | 2023 |
| 15 | 0.0250 | 2052 | 0.0275 | 2050 |
| 16 | 0.0140 | 2092 | 0.0162 | 2085 |
| 17 | 0.0068 | 2163 | 0.0082 | 2142 |
| 18 | 0.0031 | 2310 | 0.0039 | 2251 |
| 19 | 0.0014 | 2634 | 0.0017 | 2481 |

* In arbitrary units

## 4. Robustness of the results

The bell-shaped distributions obtained for the evolution of complexity in Figures 3 and 4 are similar to – have the same FWHM with – the one obtained in the old study of 2002.[Modis, 2002] The complexity of the 13$^{th}$ milestone (**Interne**/sequencing of the human genome) had been forecasted then to be at the top of the complexity curve. We see now that it has come only slightly before the top. To a large extent, the new results are compatible with those obtained in 2002, which is encouraging considering that the data have now changed in several ways: (1) we have focused on evolutionary milestones concerning only humans, (2) the accuracy of the dates and the associated uncertainties has been improved, (3) new milestones have been added the most significant one being the appearance of AI in 2023.

The reader's attention is drawn to the fact that the trends in Figures 3 and 4 remain purely exponential (straight line on the lower graph with the logarithmic vertical scale) with extremely low values for most of the range. The trend begins deviating from exponential only recently, namely from milestone No. 9 (**Renaissance**) onward. So even if we ignored several earlier milestones, we would not obtain a significantly different fit.

Moreover, Table III shows little difference in the forecast results obtained by studying all 128 milestones and by studying only the 56 major ones. This is because the dates assigned to the clusters formed by all milestones are heavily weighted by the major milestones, which are often represented by multiple entries.

### 4.1 Our process of remembering and forgetting

In 2012, the author was invited to contribute a piece to a book with title *Singularity Hypothesis: A Scientific and Philosophical Assessment*.[Edenm et al, 2012] In his contribution the author succeeded in including the following text despite the vehement objections of the editor.

"Could it be that on a large scale there may be no acceleration at all? Could it be that the crowding of milestones in Figure (…) is simply a matter of perception? The other day I was told that I should have included FaceBook as a milestone. ‘It is just as important as the Internet,’ she told me. Would Thomas Edison have thought so? Will people one thousand years from now, assuming we will survive, think so? Will they know what FaceBook was? Will they know what the Internet was?

It is natural that we are more aware of recent events than events far in the past. It is also natural that the farther in the past we search for important events the fewer of them will stick out in society’s collective memory. This by itself would suffice to explain the exponential pattern of our milestones. It could be that as importance fades with the mere distancing from the present it ‘gives the appearance,’ in John von Neumann’s words, that we are ‘approaching some essential singularity.’ But this has nothing to do with the year 2045, 2025, today, von Neumann’s time – the 1950s – or any other time in the past or the future for that matter.”

As much as there is some truth in the above reasoning, it is safe to assume that if we select a handful of milestones with only the *highest* importance over a period of 700,000 years, the importance of these milestones is likely to survive the passage of time. On the other hand, the forecasts in Table III have resulted from fits, which were influenced heavily by the six most recent milestones, i.e. from the **Renaissance** onward, as explained earlier. This relatively short historical window eliminates to a large extent the possibility that milestones of utmost importance in this period may have been forgotten.

At the same token, it is unlikely that there have been milestones in recent decades or centuries, which have not yet been recognized as such. The importance of most milestones was recognized instantly (e.g. Internet and AI,) and in some cases, the importance was recognized well before their appearance (e.g. nuclear energy.)

## 5. Discussion

Scholars relying mostly on their intuition had forecasted AI to show up during the early 21$^{st}$ century [Cerf, 2000], by 2029 [Kurzweil, 2005], or “in the next few decades” [Bostrom, 2014]. In a publication using the same approach as this paper, AI was anticipated to show up by 2033.[Modis, 2022] The fact that it came ten years earlier causes the complexity of the 13$^{th}$ milestone (Interne/sequencing of the human genome) to be higher. This is because the shorter the distance to the next milestone, the higher the complexity according to Equation (1). As it stands, the complexity added by the 14$^{th}$ milestone (AI) is now forecasted to be slightly higher than the complexity added by the 13$^{th}$ milestone (Internet/sequencing of the human genome.)

The next evolutionary milestone of comparable importance is forecasted around 2050-2052 and will add significantly less complexity than AI but more than the 12$^{th}$ milestone (nuclear energy/DNA/transistor) according to Table III. It could consist of a group of significant achievements in bioengineering, neuroscience, nanotechnology, and quantum computing clustered around that date, as we saw around the turn of the 20$^{th}$ century with modern physics (milestone No. 11.)

Following AI’s appearance complexity begins on a declining trajectory. It is a direct consequence of having described the accumulation of complexity (related to entropy) with a natural-growth curve (logistic function,) which so far seems to hold quite well.[Modis, 2022] The idea that our world’s complexity will be decreasing in the future may seem difficult to believe, but such an unimodal pattern (namely low at the beginning and the end but high in between, not unlike the normal – Gaussian – distribution) is commonplace in everyday life. It is associated with a reversal appearing at extremes. A large number of "hard" scientists, besides Carroll, Gell-Mann, and Kauffman mentioned in the introduction, have argued for a bell-shaped evolution of complexity. Here are a few of them:

In their book *Into the Cool: Energy Flow, Thermodynamics, and Life*, Eric Schneider and Dorion Sagan propose the idea of increasing and decreasing complexity in relation to entropy. They argue that in complex systems as entropy increases, there is an initial increase in complexity, but eventually the system becomes too disordered and its complexity breaks down.[Schneider and Sagan, 2005]

In dynamical systems, both periodic and random processes are considered simple, while complex and chaotic processes lie in between.[Adami, 2002]

Huberman and Hogg demonstrate that in discrete systems complexity takes low values for both ordered and disordered states while increases for intermediate states, tracing out an almost regular bell shape.[Hubermann and Hog, 1986]

The idea that complexity first increases and then decreases as entropy increases has also been advocated by two more researchers.[Grassberger, 1989] [Li, 1991]

In his book *From Eternity to Here: The Quest for the Ultimate Theory of Time*, Sean Carroll argues that complexity first increases and then decreases as entropy increases in a closed system.[Carroll, 2010] In fact, with two collaborators Carroll attempted to quantify this phenomenon for a cup of coffee with cream. In the beginning, when the cream rests calmly on top of the coffee, the entropy and the complexity of the system are small. In the end when cream and coffee are thoroughly mixed, the entropy is maximal but the complexity is small again because it is trivial to describe the system. Somewhere in the middle, while the entropy is growing, the complexity becomes maximal.[Aaronson et al, 2014] For some reason their work is yet to be published but in an archived draft, the authors claim to have obtained quantitative results demonstrating that complexity grows at first but decreases later as entropy reaches its final maximum. Such a quantitative demonstration has now been published for a simpler system, the case of throwing a fair dice very many times. As excessive wear and tear progressively morphs the dice cube into a sphere, the entropy keeps growing but the complexity first increases and eventually decreases.[Modis, 2024]

With information-related definitions for entropy and complexity, a simple mathematical relationship between them has been established, namely the latter being the time derivative of the former. It follows that if entropy traces out an S-shaped curve, complexity will trace out a bell-shaped curve.[Modis, 2022, 2024] This relationship – complexity being the derivative of entropy – cannot be rigorously generalized in all cases because for a system in equilibrium, with entropy being independent of time, it would imply no complexity whatsoever, which would be wrong. However, to the extent that all definitions for entropy (and also for complexity) are related to one another through mathematical and conceptual connections the validity and usefulness of such a relationship can be appreciated in general only qualitatively. Accordingly, the complexity of a system in equilibrium may not be equal to zero, but it is indeed very small.

Entropy whether defined as "a measure of the amount of disorder" or "the information content"[Shannon, 1948] grows monotonically and generally along an S-shaped trajectory. As entropy approaches the final maximum the information content becomes uninteresting because there is maximum disorder, everything is everywhere (100% random distribution.) Information content begins becoming uninteresting at the inflection point of entropy's trajectory.

Complexity whether defined as “how difficult it is to describe” or “the capacity to incorporate information” first grows as entropy grows but later declines tracing some sort of a bell-shaped trajectory. This would correspond to Organized Complexity as opposed to Disorganized Complexity in the distinction made by Weaver many decades ago.[Weaver, 1947] In the view of Siegenfeld and Bar-Yam we are dealing with a correlated system, where complexity gradually increases as one examines the system in greater and greater detail, i.e. smaller and smaller scale, see their Figure 2.[Siegenfeld and Bar-Yam, 2020] Aaronson et al. attribute to complexity the quality of “interestingness,” which becomes maximal when complexity goes over a maximum, halfway through the entropy growth process.[Aaronson et al, 2014]

Another grand-scale example is our solar system, which came into existence some 4.6 billion years ago. Our sun has according to current scientific understanding 5 billion years of hydrogen fuel left before it begins to run out and enter its red-giant phase. The entropy (information content) of our solar system increased rapidly in the beginning but toward the end it will slowly (asymptotically) reach a maximum, which will be full of uninteresting information. Meanwhile, the complexity increased as entropy increased in the beginning, and will decrease at the end. It will go over a maximum roughly halfway between the beginning and the end, and not far from our times (in cosmic timescale.) At this time entropy is being generated at a maximum rate. No significant increase in the complexity of our solar system should be expected in the future. Equating complexity with interest argues that we are traversing the most interesting times of our solar system!

On another front, complexity seems to be linked to the population in intricate ways. In the next two sections we see first, that people belonging to the ill-understood baby boom have life spans that straddle the complexity peak. And second, that complexity is most likely modulated by the *rate* of growth of the population rather than its actual size.

### 5.1 The baby boom

The baby boom has been observed in America but also in many other parts of the world. The popular explanation, namely that soldiers came back from the war and indulged in making babies, is fallacious. We see in the graph at the top of Figure 5 that the downward trend of annual live births beginning in the early 20th century can be well-described by a downward-pointing logistic curve (thick gray line,) which is fitted on the data of the periods: 1909-1933 and 1973-2006. Deviations from the curve begin as early as 1934 and extend to 1972. The effect of soldiers going to and coming back from the war is indeed visible but only as a small wiggle between 1942 and 1948, a much smaller effect than the overall process.

Isolating the baby-boom data by subtracting the trend from the data numbers yields a bell-shaped distribution, see lower graph on Figure 5, itself well-described by the rate of growth (the derivative) of another logistic function, fitted over the same period, see the thick gray line in the lower graph of the figure. The goodness of the two fits can be appreciated visually but also by the Graphical Analysis of Residuals mentioned earlier, see Table IV.

## Table IV - Fit results for the two baby-boom logistics

| **Functions fitted** | | | | | **Goodness of Fit** | | |
|---|---|---|---|---|---|---|---|
| $C - \frac{M}{(1+e^{-\alpha(x-x_0)})}$ | $\alpha$ | $M$ | $x_0$ | $C$ | $R^2$ | *Slope* | *Intercept* |
| | 0.205 | 15.23 | 1927.7 | 30.29 | 0.986 | 1.000 | -0.0002 |
| $\frac{M\alpha}{(1+e^{-\alpha(x-x_0)})(1+e^{\alpha(x-x_0)})}$ | $\alpha$ | $M$ | $x_0$ | | $R^2$ | *Slope* | *Intercept* |
| | 0.176 | 248.6 | 1953.4 | | 0.947 | 1.053 | -0.393 |

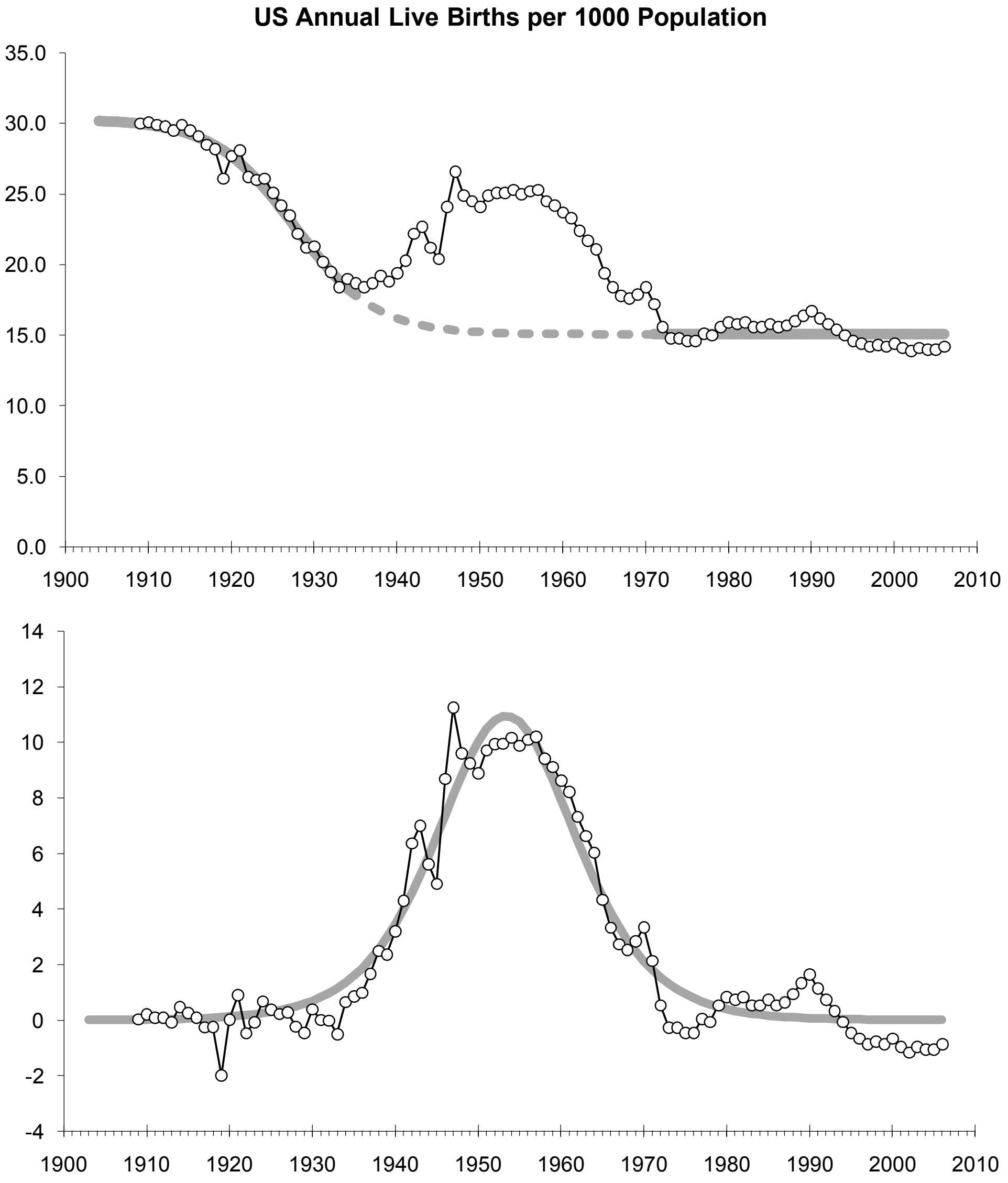

Data source: U.S. Census Bureau, Statistical Abstract of the United States: 2003

Figure 5. A declining logistic-growth curve (thick gray line on the upper graph) permits extraction of the baby-boom data, open circles on lower graph, which are well described by the rate of growth of another logistic function (thick gray line on the lower graph.)

Assuming a life expectancy of 80 years the span of baby boomers extends from 1934 to 1972+80=2052, and coincides squarely with the peak of the complexity curve, see delimitations by the dotted lines in Figure 6.

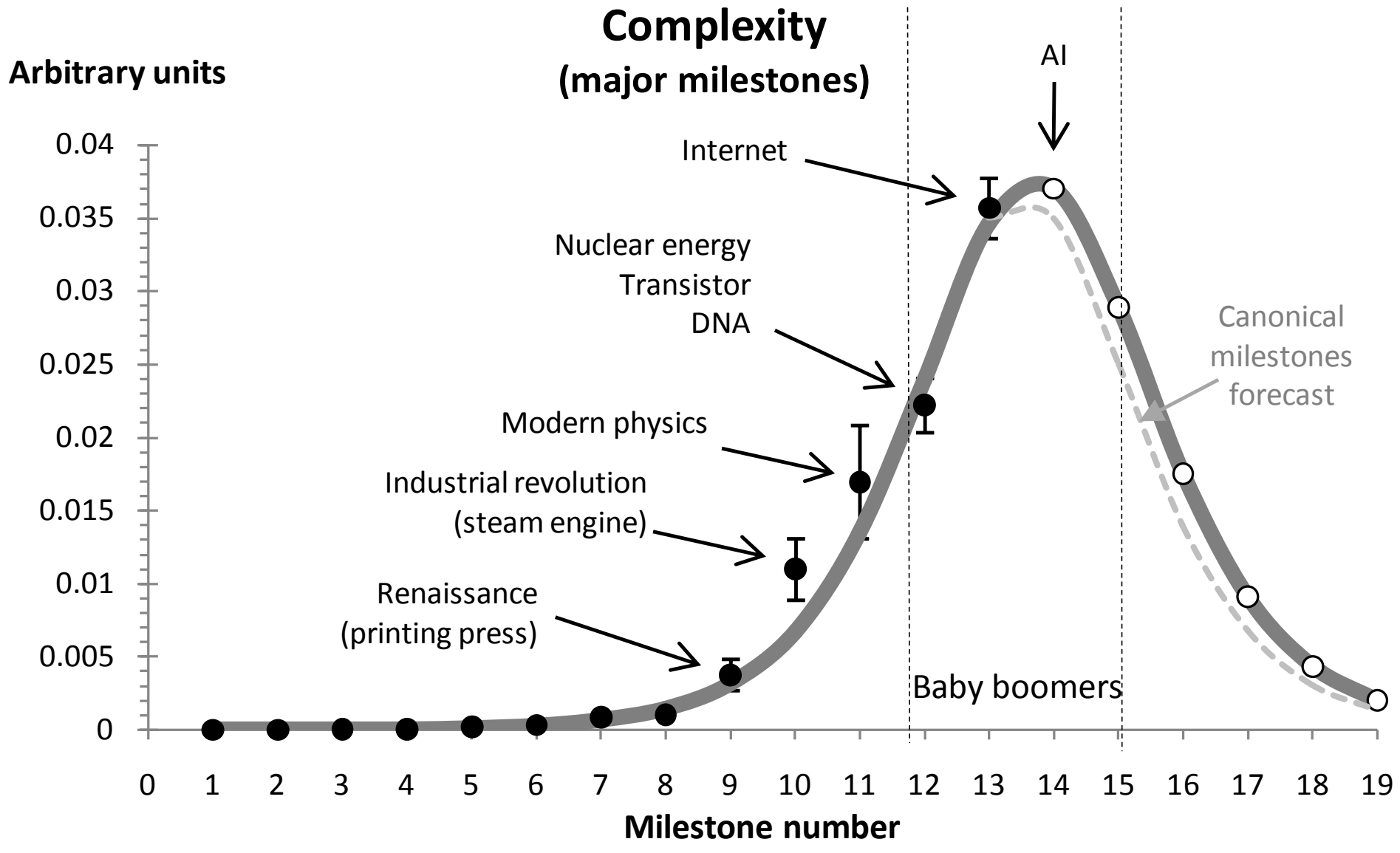

Figure 6. This is the same as the top graph in Figures 2 and 3 but with the baby-boom-generation span superimposed (delimited by the two dotted vertical lines), and some annotations.

### 5.2 The world population

There is one more phenomenon that "resonates" with the bell-shaped complexity curve determined earlier: the growth of the world population. In Figure 7 we see the evolution of the world population since 1950, a period during which its growth has been most dramatic and well documented. There is excellent agreement between the data and the fitted logistic growth curve (gray line.) In fact, the forecast for 2040 of 9.0 billion agrees with the forecast from the United States Census Bureau of 9.17 billion. It is superfluous at this point to give quantitative goodness-of-fit parameters.

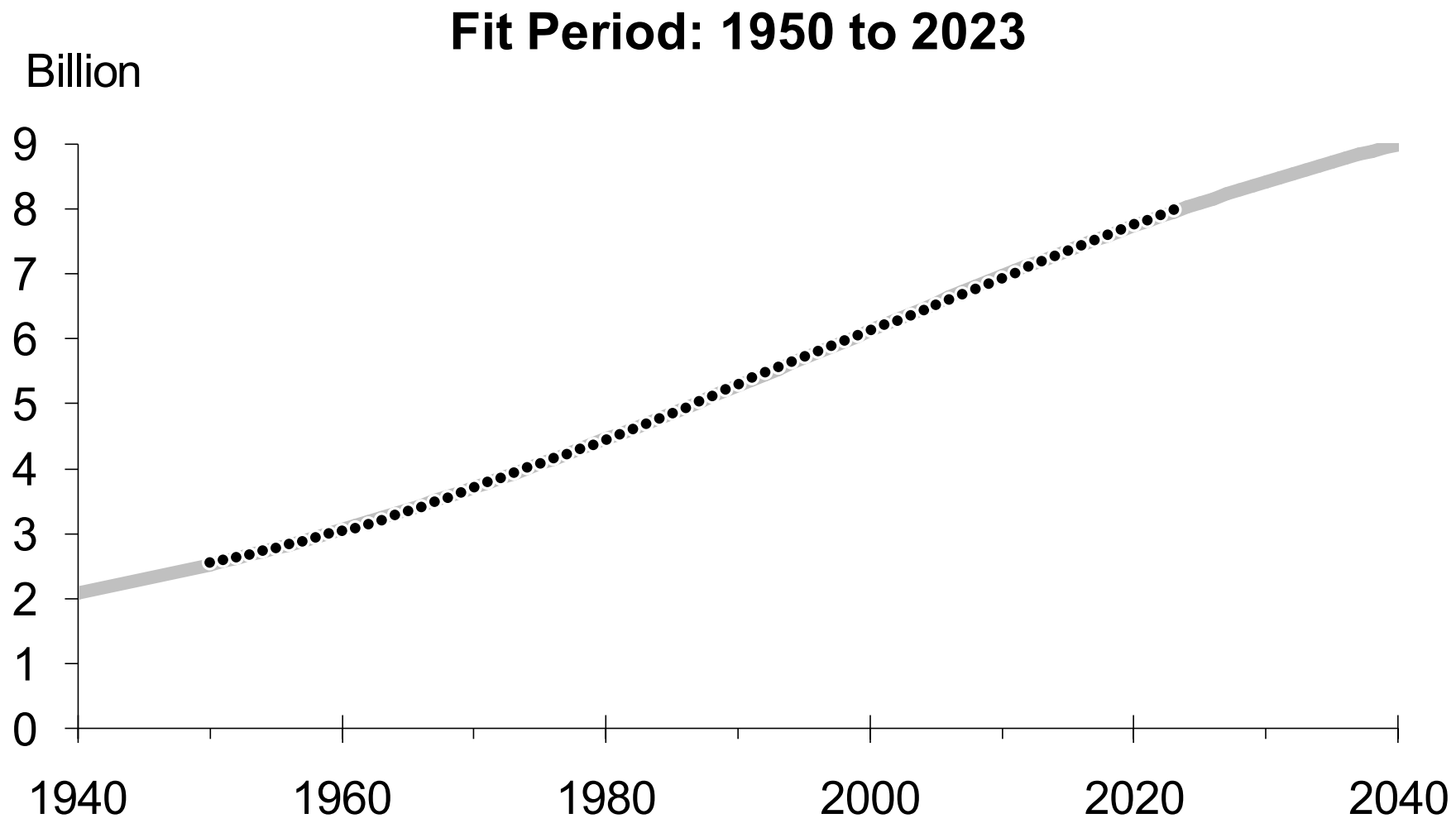

Data source: International Database U.S. Census Bureau 2024

Figure 7. The gray line is a logistic fit to the world population data (black dots.)

We can now compare in Figure 8 the world population's rate of growth (black line) with the complexity bell-shape curve (gray line), as previously determined in Figure 4, now expressed as a function of time using the values from Table III. The comparison is interesting. The population bump precedes the complexity bump by around 25 years. Also interesting is the fact that the baby-boom-generation span straddles these curves.

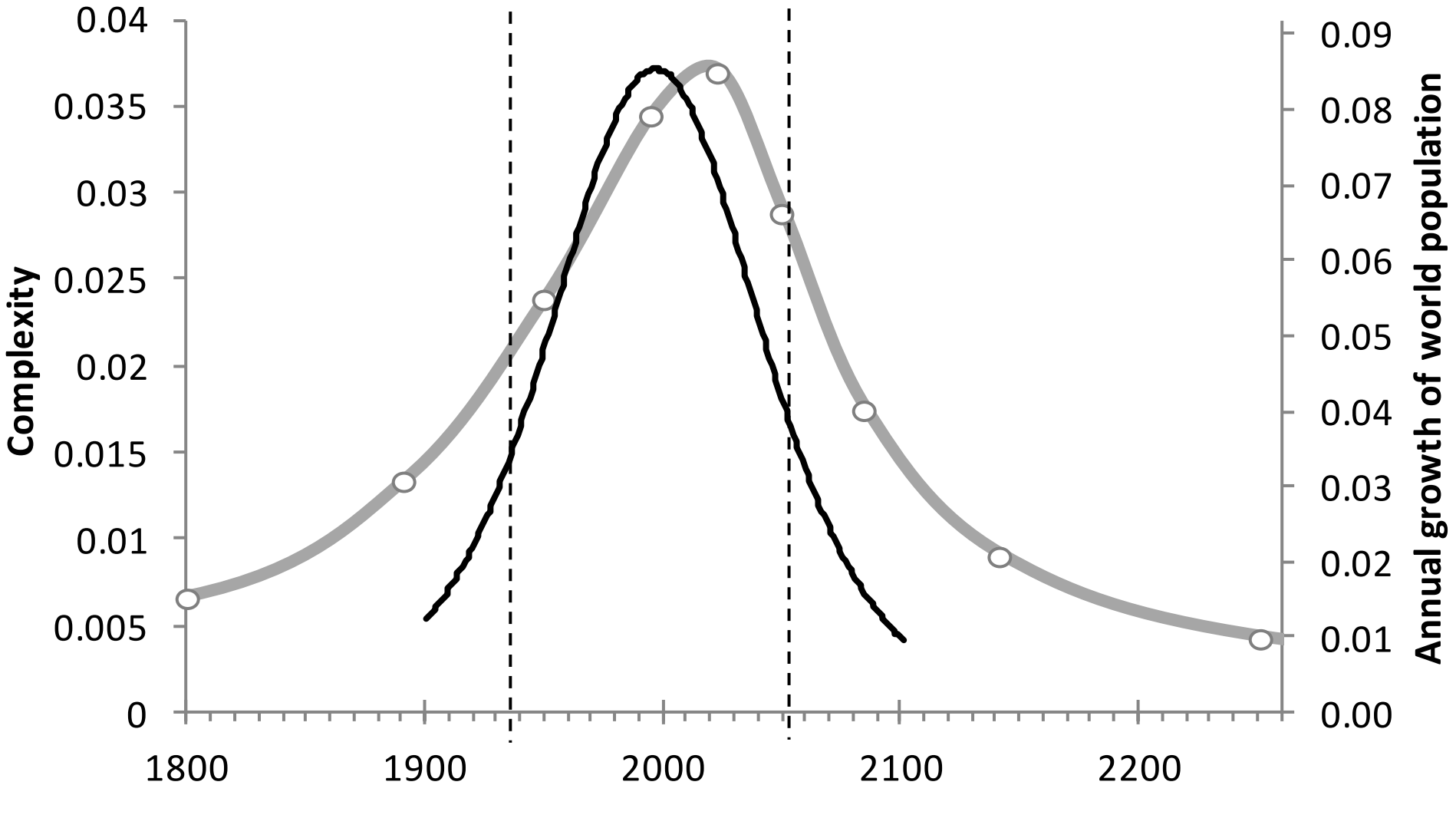

Figure 8. The gray line of complexity (read on the left vertical axis) is the same as the gray line in Figure 4. The black line (read on the right vertical axis in billions) is the rate of growth of the logistic function indicated by the gray line in Figure 7. The baby-boom-generation span is again delimited by the two dotted vertical lines.

It must be pointed out that the population and the baby boom phenomena have very different sizes. The world population increased by 1.9 billion between 1934 and 1971 while the total live births of baby boomers for this period amounts to 234.6 million, a ratio of 0.00123. Granted, we are considering only US baby boomers here, but even if they represent only 10-15% of the world's total baby boomers, the baby boom phenomenon still amounts to a minuscule perturbation on the evolution of the world population.

## 6. Conclusions

The complexity curve of the human system has grown following a bell-shaped curve, has gone over a peak, and now is about to begin decreasing. The 2002 work [Modis,2002] concluded that "we are sitting on top of the world." Following the emergence of AI this conclusion seems now to be reinforced. The next evolutionary milestone of comparable importance is expected around 2050-2052 and should add less complexity than AI but more than milestone No. 12 grouping together nuclear energy, DNA, and the transistor.

The lifespan of baby boomers coincides squarely with the complexity peak as if, through some kind of serendipity, an enhanced number of people were meant to live through these excessively complex years. The baby-boom generation will have witnessed more complexity during their lives than anyone else before or after them. Inversely, baby boomers could be considered responsible for contributing an excess complexity to the human system. After all, the development of the Internet and AI took place mostly during the prime of the baby boomers' active lifespan. But such contribution would be limited due to the relatively small number of baby boomers compared to the overall world population as mentioned earlier.

More significantly, the rate of growth of the world population has followed a bell-shaped trajectory going over a maximum in 1997. But already from the 1980s onward this population curve preceded the complexity curve by 23 to 25 years, which is about the time it takes a young man or woman before they are able to add complexity to the human system in a significant way. These observations do not constitute proof that the population's *rate* of growth dictates how complexity will evolve. However, there is an argument that can be made in that direction. An increasing population increases the entropy of the human system, and complexity follows the rate of growth of entropy, at least qualitatively.[Modis, 2022, 2024] Consequently, the rate of growth of the population could reasonably dictate how complexity will evolve. Economist Robin Hanson argues "Population decline implies innovation decline." If we assume that innovation and complexity are intimately linked, we can specify Hanson's saying as "A decline in population's *rate* of growth implies innovation decline 25 years later."

Future milestones will continue arriving, albeit at increasingly longer time intervals, so the end of the world is certainly not around the corner. Still, one thing is clear. Humans and life in general have demonstrated that they evolve and thrive when complexity is increasing. They will probably do less well when it is decreasing.

It is in society's interest to flatten the complexity bell-shaped curve to whatever extent this is possible. It implies slowing down complexity's rate of change (i.e. decrease the parameter $\boldsymbol{\alpha}$ in Tables I and II.) A study has established correlations between the three parameters of the logistic function. In particular, a negative correlation was found between the level of the final ceiling $\boldsymbol{M}$ (the niche capacity) and the rate of change $\boldsymbol{\alpha}$ (the slope) of the logistic.[Debecker and Modis, 1994] This study was revisited when the world rushed toward flattening the curve of the COVID-19 pandemic, pointing out that flattening the curve would increase the total number of victims.[Debecker and Modis, 2021]

By lowering complexity's rate of change – it could be done by simply embracing slowing down or doing-less practices – people will not only enjoy complexity for a longer time, but they will also achieve a bigger overall cumulative complexity, and consequently entropy, both of which are ingredients indispensable for their well-being, as mentioned at the beginning of the Introduction. Movements – such as Minimalism, Slow Living, and Degrowth, which are emerging more and more often in the 21st century, could be evidence of society's unwitting attempts toward a beneficial flattening of complexity's bell-shaped curve.

## Appendix A – The Milestones Data

The dates of the 14 milestones below generally represent weighted averages of clustered events (see Figure 1) not all of which are mentioned in this table. That is why some events may appear dated somewhat off, e.g., WWI. Highlighted in bold are the most outstanding events in the cluster, i.e. *major* milestones as defined in the text, see Figure 2.

| No. | Milestone | Years Before 2000 |
|---|---|---|
| 1. | **Domestication of fire** | **700,000** |
| 2. | **Emergence of *Homo sapiens*,** acquisition of spoken language | 266,667 (**300,000**) |
| 3. | **Earliest burial of the dead** | **100,000** |
| 4. | **Rock art,** Seafarers settle in Australia | 46,250 (**40,000**) |
| 5. | **First cities, invention of agriculture** | **10,071** |
| 6. | **Development of the wheel, writing,** bronze metallurgy, Giza pyramid, Egyptian Kingdoms, Mycenaean, Olmec culture, Hammurabic legal codes in Babylon | 4,649 (**5450**) |
| 7. | **Buddha, democracy, city states,** the Greeks, Euclidean geometry, Archimedean physics, Roman Empire, iron metallurgy, astronomy, Asokan India, Ch'in Dynasty China | 2,478 (**2536**) |
| 8. | **Number system notation, zero and decimals,** invention of the compass, Rome falls | 1,338 (**1415**) |
| 9. | **Renaissance** (printing press), **experimental method in science** (Galileo, Kepler), gunpowder, discovery of New World | 542 (**487**) |
| 10. | **Industrial revolution** (steam engine), French/American revolutions, Enlightenment era | 231 (**200**) |
| 11. | **Modern physics** (widespread development of science and technology: electricity, radio, telephone, television, automobile, airplane), Einstein, WWI, Globalization | 104 (**97**) |
| 12. | **Nuclear energy, DNA structure described, transistor**, Sputnik, WWII, Cold War, establishment of the United Nations | 49 (**51**) |
| 13. | **Internet,** sequencing of the human genome | **5** |
| 14. | **Artificial Intelligence,** big data, social media | **-23** |

**Acknowledgments**

The author would like to thank Alain Debecker, Athanasios G. Konstantopoulos, Pierre Darriulat, Yorgo Modis, and the Nobel laureates Sir John Walker and Sir Peter Ratcliffe for useful discussions.

During the preparation of this work, the author used ChatGPT in order to distill a set of 25 milestones (breaks in historical perspective.) After using this tool, the author carefully reviewed and edited the content and takes full responsibility for its validity.

**Conflict of interest statement**

There is no conflict of interest.

**Funding statement**

There has been no funding for this work.

**Data availability statement**

The data used for this analysis are listed in Appendix A. Publically available data, whenever used, are specified in the text.

## Biographical Note

Theodore Modis is a physicist, engineer, futurist, and international consultant. He is the author/co-author of over one hundred articles in scientific and business journals and ten books. He has on occasion taught at Columbia University, the University of Geneva, at business schools INSEAD and IMD, and at the leadership school DUXX, in Monterrey, Mexico. He has been on the advisory board of the journal Technological Forecasting & Social Change since 1991. He is the founder of Growth Dynamics, an organization specializing in strategic forecasting and management consulting: http://www.growth-dynamics.com